\documentclass[12pt,thmsa]{article}
\usepackage{amsmath}
\usepackage{amsfonts}

\input{tcilatex}
\begin{document}

\title{Charge without charge, regular spherically symmetric solutions and
the Einstein-Born-Infeld theory}
\author{D.J. Cirilo Lombardo \\
%EndAName
Bogoliubov Laboratory of Theoretical Physics \\
Joint Institute of Nuclear Research, \\
Dubna, Moscow Region, 141980, Russia\thanks{%
e-mails:diego@thsun1.jinr.ru diego77jcl@yahoo.com.}}
\maketitle

\begin{abstract}
The aim of this paper is to continue the research of JMP 46, 042501 (2005)
of regular static spherically symmetric spacetimes in Einstein-Born-Infeld
theories from the point of view of the spacetime geometry and the
electromagnetic structure. The energy conditions, geodesic completeness and
the main features of the horizons of this spacetime are explicitly shown. A
new static spherically symmetric dyonic solution in Einstein-Born-Infeld\
theory with similar good properties as in the regular pure electric and
magnetic cases of \ our previous work, is presented and analyzed. Also, the
circumvention of a version of "no go" theorem claiming the non existence of
regular electric black holes and other electromagnetic static spherically
configurations with regular center is explained by dealing with a more
general statement of the problem.
\end{abstract}

\section{\protect\bigskip The regularity problem}

Last years considerable interest in regular black hole solutions has been
renewed$^{6}$, and in particular, to those based in nonlinear
electrodynamics (NED)$^{7,8}$. In our previous work$^{1}$we presented a new 
\textit{exact} spherically symmetric solution of the Einstein-Born-Infeld \
(EBI) equations. The metric is \textit{regular} everywhere in the sense that
was given by B. Hoffmann and L. Infeld$^{4}$ in 1937 and, when the intrinsic
mass of the system is zero, the EBI\ theory leads to identification of the
gravitational with the electromagnetic mass. Explicitly the line element
that we obtained in ref.$^{1}$ is 
\begin{equation}
ds^{2}=-e^{2\Lambda }dt^{2}+e^{2\mathcal{F}\left( r\right) }\left[
e^{-2\Lambda }\left( 1+r\,\ \partial _{r}\mathcal{F}\left( r\right)
\,\right) ^{2}dr^{2}+r^{2}\left( d\theta ^{2}+\sin ^{2}\theta \,d\varphi
^{2}\right) \right] ,  \tag{1}
\end{equation}%
in particular the $g_{tt}$ coefficient, takes the following form : 
\begin{equation}
e^{2\Lambda }=1-\frac{2M}{Y}-\frac{2b^{2}r_{o}^{4}}{3\left( \sqrt{%
Y^{4}+r_{o}^{4}}+Y^{2}\right) }-\frac{4}{3}b^{2}r_{o}^{2}\,_{2}F_{1}\left[
1/4,1/2,5/4;-\left( \frac{Y}{r_{0}}\right) ^{4}\right]  \tag{2}
\end{equation}%
here $M$ is an integration constant, which can be interpreted as an
intrinsic mass, and $_{2}F_{1}$ is the Gauss hypergeometric function$^{16}.$
Specifically, for the form of the $\mathcal{F}\left( r\right) $ defined by
the expression (1), the coordinate function $Y(r)$is related with the
function $\mathcal{F}\left( r\right) $ as 
\begin{equation}
Y\left( r\right) ^{2}\equiv \left[ 1-\left( \frac{r_{o}}{a\left\vert
r\right\vert }\right) ^{n}\right] ^{2m}r^{2}=e^{2\mathcal{F}\left( r\right)
}r^{2}  \tag{2.1}
\end{equation}%
Regarding the historical origin of the question, in 1937 B. Hoffmann and L.
Infeld$^{4}$ introduce a regularity condition on the new field theory of M.
Born$^{10}$with the main idea of to solve the lack of uniqueness of the
function action\footnote[1]{%
The new field theory initiated in 1934 by M. Born$^{10}$ introduces in the
classical equations of the electromagnetic field a characteristic length $%
r_{0}$ representing the radius of the elementary particle through the
relation 
\begin{equation*}
r_{0}=\sqrt{\frac{e}{b}}
\end{equation*}%
where $e$ is the elementary charge and $b$ the fundamental field strength
entering in a non-linear Lagrangian function. It was originally thought that
the Lagrangian $\mathcal{L}_{BI}=\sqrt{-g}L_{BI}=\frac{b^{2}}{4\pi }\left\{ 
\sqrt{-g}-\sqrt{\left\vert \det (g_{\mu \nu }+b^{-1}F_{\mu \nu })\right\vert 
}\right\} $ was the simplest choice which would lead to a finite energy for
an electric particle. This is, however, not the case. It is possible to find
an infinite number of quite different action functions, each giving simple
algebraic relations between the fields and each leading to a finite energy
for an electric particle.}. They have already seen that the condition of
regularity of the field gives the restriction in the spherically symmetric
electrostatic case $E_{r}=0$ for $r=0$.

In the general theory they applied the regularity condition not only to the $%
F_{\mu \nu }$ field but also to the $g_{\mu \nu }$ field. The regularity
condition for the general theory was that:

\textit{Only those solutions of the field equations may have physical
meaning for which space-time is everywhere regular and for which the }$%
F_{\mu \nu }$\textit{\ and the }$g_{\mu \nu }$\textit{\ fields and those of
their derivatives which enter in the field equations and the conservation
laws exist everywhere.}

In the general relativity form of the original new field theory the
requirement that there be no infinities in the $g_{\mu \nu }$ forces the
identification of gravitational with electromagnetic mass. In$^{4}$ B.
Hoffmann and L. Infeld have used for such identification the line element of
the well known monopole solution studied by B. Hoffmann$^{3}$ in 1935 
\begin{equation*}
A\left( r\right) \equiv 1-\frac{8\pi G}{r}\int_{0}^{r}\left[ (r^{\prime
4}+1)^{1/2}-r^{^{\prime }2}\right] dr^{\prime }
\end{equation*}%
that is originated by an EBI action as in equation (1) of $^{1}$ (see
footnote 1). This line element approximates the Schwarzschild form for $r$
greater than the electronic radius but avoid the infinities of that line
element for $r=0$. However is still a singularity of conical type at the
origin. When $r\rightarrow 0$ the above expression for \ $A$, gives 
\begin{equation*}
A\rightarrow (1-8\pi G)\equiv \beta
\end{equation*}%
so $ds^{2}$ becomes 
\begin{equation*}
ds^{2}=-\beta dt^{2}+\beta ^{-1}dr^{2}+r^{2}\left( d\theta ^{2}+\sin
^{2}\theta \,d\varphi ^{2}\right)
\end{equation*}%
Therefore the origin (it is, at $r=0$) is a conical point and not regular.
Note that, because the conical point, no coordinate can be introduced which
will be non singular at $r=0$ and derivatives are actually undefined at this
point.

This problem with the conical singularities at $r=0$ , that destroy the
regularity condition, makes that in the reference$^{4}$ B. Hoffmann and L.
Infeld change the action of the Born-Infeld form for other non-linear
Lagrangian of logarithmic type, were the metric takes the form :%
\begin{equation*}
A\left( r\right) \equiv 1-\frac{8\pi G}{r}\int_{0}^{r}r^{\prime 2}Log\left[ 
\frac{r^{\prime 2}}{r^{\prime 4}+1}\right] dr^{\prime }
\end{equation*}%
The new logarithmic action does not presented such difficulties at $r=0$,
but only for the electric field the solution is allowed: there are not
possibility of magnetic monopoles. Also, the $\mathbb{F}^{\mu \nu }\equiv 
\frac{\partial L}{\partial F_{^{\mu \nu }}}$ field has a divergent behaviour
at $r\rightarrow 0$. Therefore, the only requirement of regularity in order
to ruled out a specific form of the Lagrangian seems to be \textit{%
insufficient}. Moreover: \ in ref. $^{1}$we was able to solve the problem of
regularity with the original (first principles) BI Lagrangian, that permits
a symmetric treatment of magnetic and electric field configurations with a
full generalized duality (as was also shown by the authors in$^{12}$).

Returning to the summary of the main features of the line element (1), a
regular, asymptotically flat solution with the electric field and
energy-momentum tensor both regular, in the sense of B. Hoffmann and L.
Infeld is when the exponent numbers of $Y(r)$ take the following particular
values: 
\begin{equation*}
n=3\ \ \ and\ \ \ \ \ m=1
\end{equation*}%
and the integration constant $M$ can have any value, not necessarily zero
(this fact what not advertised by the authors in $^{1}$. We will return to
this point in Section 8). The spacetime is singularity free and geodesically
complete. In this case and for $r>>\frac{r_{0}}{a},$ we have the following
asymptotic behaviour for $Y\left( r\right) $ and $-g_{tt}\,$, that does not
depend on the $a$ parameter$^{{}}$%
\begin{equation*}
Y\left( r\right) \rightarrow r\ \text{ \ \ \ \ \ }\left( r>>\frac{r_{0}}{a}%
\right)
\end{equation*}%
\begin{equation*}
e^{2\Lambda }\simeq 1-\frac{2M}{r}-\frac{8b^{2}r_{o}^{4}K\left( 1/2\right) }{%
3r_{o}r}+\frac{2b^{2}r_{o}^{4}}{r^{2}}+...
\end{equation*}%
As we pointed out in ref.$^{1}$, a distant observer will associate with this
solution a total mass 
\begin{equation*}
M_{eff}=M+\frac{4b^{2}r_{o}^{4}K\left( 1/2\right) }{3r_{o}}
\end{equation*}%
total charge 
\begin{equation*}
Q^{2}=2b^{2}r_{o}^{2}
\end{equation*}%
being the electromagnetic mass 
\begin{equation}
M_{el}=\frac{4b^{2}r_{o}^{4}K\left( 1/2\right) }{3r_{o}}  \tag{2.2}
\end{equation}%
Notice that the $M_{el}$ is necessarily positive, which was not the case in
the Schwarzschild line element where $M$ is an integration constant. The
important reason for to take the constant $M=0$ is that we must regard the
quantity (let us to restore by one moment the gravitational constant $G$) 
\begin{equation*}
4\pi G\int_{Y(r=0)}^{Y(r)}T_{0}^{0}\left( Y\right) Y^{2}dY
\end{equation*}%
as the \textit{gravitational mass} causing the field at coordinate distance
r from the pole. In our case $T_{0}^{0}$ is given by $-\frac{b^{2}}{4\pi }%
\left( 1-\sqrt{\left( \frac{r_{0}}{e^{G}}\right) ^{4}+1}\right) $. This
quantity is precisely $M_{el}$ (in gravitational units) given by (2.2), the 
\textit{total electromagnetic mass} within the sphere having its center at $%
r=0$ and coordinate $r.$ We will take $M=0$ in the rest of the analysis
being consistent with the gravitational-electromagnetic identification.

On the other hand, the function $Y\left( r\right) $ for the values of the $m$
and $n$ parameters given above has the following behaviour near of the
origin 
\begin{equation*}
\text{for\ }a<0\text{ \ \ \ \ \ \ \ \ when\ }r\rightarrow 0,\ Y\left(
r\right) \rightarrow \infty
\end{equation*}%
\begin{equation*}
\text{for\ }a>0\text{ \ \ \ \ \ \ \ \ \ when \ }r\rightarrow 0,\ Y\left(
r\right) \rightarrow -\infty
\end{equation*}%
Notice that the case \ $a>0$ will be excluded because in any value $%
r_{0}\rightarrow $ $Y\left( r_{0}\right) =0$, the electric field takes the
limit value $b$ and the condition $\left. F_{01}\right\vert _{r=r_{o}}<b$ is
violated (and the $T_{00}$ diverges). For $M=0$ and $a<0,$expanding the
hypergeometric function, we can see that the $-g_{tt}$ coefficient has the
following behaviour near the origin 
\begin{equation*}
e^{2\Lambda }\simeq 1-\frac{8b^{2}r_{o}^{4}K\left( 1/2\right) }{3r_{o}}%
r^{2}\left( \frac{\left\vert a\right\vert }{r_{0}}\right)
^{3}+2b^{2}r_{0}^{4}\ r^{4}\left( \frac{\left\vert a\right\vert }{r_{0}}%
\right) ^{6}+...
\end{equation*}%
The metric and the energy-momentum tensor remain \textit{both }regulars at
the origin (it is: $g_{tt}\rightarrow -1,T_{\mu \nu }\rightarrow 0$ $\ $\
for $r\rightarrow 0$). It is not very difficult to check that (for $m=1$ and 
$n=3$) the maximum of the electric field (see figures) is not in $r=0$ , but
in the \textit{physical} \textit{border} of the spherical configuration
source of the electromagnetic fields (this point is located around $%
r_{B}=2^{1/3}\frac{r_{0}}{\left\vert a\right\vert }$). It means that $%
Y\left( r\right) $ maps correctly the internal structure of the source in
the similarly form that the quasiglobal coordinate of the reference$^{17}$
for the global monopole in general relativity but, in contrast to the case
pointed out in ref.$^{17}$, $Y\left( r\right) $ is \textit{function} from
the strict mathematical point of view. The lack of the conical singularities
at the origin is because the very well description of the manifold in the
neighborhood of $r=0$ given by the function $Y\left( r\right) .$

Because $g_{tt}$ is regular ($g_{tt}=-1,$ at $r=0$ and at $r=\infty $), its
derivative must change sign. In the usual gravitational theory of general
relativity the derivative of $g_{tt}$ is proportional to the gravitational
force which would act on a test particle in the Newtonian approximation. In
Einstein-Born-Infeld theory with this new static solution, it is interesting
to note that although this force is attractive for distances of the order $%
r_{0}<<r$ , it is actually a repulsion for very small $r$. For $r$ greater
than $r_{0},$ the line element closely approximates to the Schwarzschild
form. Thus the regularity condition shows that the electromagnetic and
gravitational mass are the same and, as in the Newtonian theory, we now have
the result that the attraction is zero in the center of the spherical
configuration source of the electromagnetic field.

The aim of this paper is to make complete the research of regular static
spherically symmetric spacetimes in EBI theories of ref.$^{1}$ emphasizing
the following points:

$\circ $ the analysis of the monopole solution presented in the previous
work from the physical and topological point of view,

$\circ $ how a version of no go theorem$^{7,8}$ claiming the non existence
of regular electric charged black holes and other electric configurations
with regular center is circumvented by dealing with a correct statement of
the problem. In fact was clearly demonstrated in$^{1}$ and results above,
that the Born-Infeld theory coupled with general relativity (EBI) lead
regular solutions where the main assumptions to have account were:

i) most general spherically symmetric line element as starting point,
compatible with the symmetry of the an anisotropic fluid$^{15}$ (the
Born-Infeld energy-momentum tensor has this form)

ii) strict physical requirements of regularity for the electric field 
\begin{equation*}
\left. F_{01}\right\vert _{r=r_{o}}<b
\end{equation*}%
\begin{equation*}
\left. F_{01}\right\vert _{r=0}=0
\end{equation*}%
\begin{equation*}
\,\ \ \ \ \ \ \ \ \ \ \ \ \ \ \ \ \ \ \ \ \ \ \ \ \ \ \left.
F_{01}\right\vert _{r\rightarrow \infty }=0\,\ \ \ \ \ \ \ \ \ \ \ \ \text{%
asymptotically \ Coulomb}
\end{equation*}%
that fix the form and behaviour of function $\mathcal{F}\left( r\right) $
being as (2.1) where $a$ is an arbitrary constant, and the exponents $n$ and 
$m$ will obey the following relation 
\begin{equation*}
mn>1\,\ \ \ \ \left( m,n\in \mathbb{N}\right)
\end{equation*}%
with 
\begin{equation*}
0<a<1\ \text{or}\ -1<a<0\ \ \ 
\end{equation*}%
depending on $m\left( n\right) $ is even or odd and 
\begin{equation*}
a\neq 0
\end{equation*}%
that put in sure and \textit{guarantees} a consistent regularization
condition not only for the electric (magnetic) field but for the
energy-momentum tensor and the line element also.

$\circ $ to show that the electric solution of $^{1}$ satisfy the weak,
dominant and strong energy conditions (WEC, DEC, SEC) and the Lagrangian has
correct Maxwellian limit at $r\rightarrow 0$ without branches.

$\circ $ a new static spherically symmetric dyonic solution in EBI\ theory
with similar good properties as for the regular pure electric and magnetic
cases, is presented and analyzed.

The organization of the paper is as follows: Section 2 is devoted to show
that the solution of$^{1}$ satisfy all the energy conditions. In Section 3
the horizons structure of the singularity free and geodesically complete
spacetime of $^{1}$ is fully analyzed and some physical aspects concerning
to the Born-Infeld radius$^{2}$ $r_{o}$, the absolute field$^{2}$ $b$ and
the cosmological constant$^{11}$ $\Lambda $ are briefly discussed. In\
Section 4 we explain with some detail how a version of no go theorem
claiming the non existence of regular electric charged black holes and other
electric configurations with regular center is circumvented.

Section 5, 6 and 7 are devoted to the new regular spherically symmetric
dyonic solution in EBI: statement of the problem, determination of the
electromagnetic structure of this spherically symmetric configuration: field
equations vs. energy-momentum conservation and aspects of the
electromagnetic field of the dyon in comparison with the pure electric and
magnetic regular cases of our previous reference $^{1}$. Finally, in Section
8 a resume of the main results and a discussion about the regular
spherically symmetric solutions in NED are given in the light of the new
results and analysis presented here. The conventions are the same that in $%
^{1}$, unless that we indicate the contrary.

\section{Energy conditions}

Is clearly important, in order to begin the study of any spacetime solution,
consider firstly if violation of the called$^{13}$ energy conditions
certainly exists. The first step to begin the analysis of the monopole
solution of $^{1}$ is to remember that the (adimensionalized) $\overline{Y}%
^{2}\left( r\right) $ function 
\begin{equation}
\overline{Y\left( r\right) }^{2}\equiv \left( \frac{Y}{r_{o}}\right) ^{2}=%
\left[ 1-\left( \frac{r_{o}}{a\left\vert r\right\vert }\right) ^{n}\right]
^{2m}\left( \frac{r}{r_{o}}\right) ^{2}  \tag{3}
\end{equation}%
with $a=-0,9$ and $m=1$ and $n=3$, never is zero with its range being%
\begin{equation}
\left. \overline{Y}^{2}\right\vert _{min}\leq \overline{Y}^{2}<\infty 
\tag{4}
\end{equation}%
\begin{equation*}
\left. \overline{Y}^{2}\right\vert _{min}=2.09987\text{ corresponding to }%
\overline{r}_{min}=1.39991
\end{equation*}%
that is the point where the electric field and the energy-momentum tensor
have the maximum value and obviously, the metric present its maximum
curvature (from here, the dependence of $Y$ on $r$ is understood) . All the
quantities of our solution presented in $^{1}$ can be written in a more
compact form through this function. Now, we put all the components of the
energy momentum tensor in the orthonormal frame (that coincide with the
energy density $\rho $ and the pressures $p_{k}$) as a functions of $%
\overline{Y}^{2}\left( r\right) $, in order to easily analyze if the
solution fulfill the energy conditions: 
\begin{equation}
\rho =T_{00}=\frac{b^{2}}{4\pi }\left( \sqrt{\left( \overline{Y}\right)
^{-4}+1}-1\right)  \tag{5a}
\end{equation}%
\begin{equation}
p_{rad}=-\rho =T_{11}=\frac{b^{2}}{4\pi }\left( 1-\sqrt{\left( \overline{Y}%
\right) ^{-4}+1}\right)  \tag{5b}
\end{equation}%
\begin{equation}
p_{\perp }=T_{22}=T_{33}=\frac{b^{2}}{4\pi }\left( 1-\frac{1}{\sqrt{\left( 
\overline{Y}\right) ^{-4}+1}}\right)  \tag{5c}
\end{equation}%
Notice that all the components of the energy momentum tensor are finite in
the spacetime of the electric nonlinear monopole.

a) The weak energy condition (WEC): 
\begin{equation*}
T_{\mu \nu }\xi ^{\mu }\xi ^{\nu }\geq 0\ (\xi ^{\mu }:\text{any timelike
vector})\Rightarrow \rho \geq 0,\ \ \rho +p_{k}\geq 0\ \ ,\ \ (k=1,2,3)
\end{equation*}%
guarantees that the energy density as measured by any local observer is
non-negative. For our case having account in expressions (4) and (5) : 
\begin{equation*}
\rho =T_{00}=\frac{b^{2}}{4\pi }\left( \sqrt{\left( \overline{Y}\right)
^{-4}+1}-1\right) \geq 0
\end{equation*}%
\begin{equation*}
\rho +p_{rad}=0
\end{equation*}%
\begin{equation*}
\rho +p_{\perp }=\frac{b^{2}}{4\pi }\left( \sqrt{\left( \overline{Y}\right)
^{-4}+1}-\frac{1}{\sqrt{\left( \overline{Y}\right) ^{-4}+1}}\right) \geq 0
\end{equation*}%
Then, the WEC\ is satisfied in all the manifold.

b) The dominant energy condition (DEC):

includes WEC and requires each each principal pressure never exceeds the
energy density which guarantees that the speed of sound cannot exceed the
light velocity c%
\begin{equation*}
T^{00}\geq \left\vert T^{ik}\ \right\vert \text{ },\ (i,k=1,2,3)\Rightarrow
\rho \geq 0,\ \ \rho +p_{k}\geq 0
\end{equation*}%
were probed before in a) and, from expressions (4) and (5)%
\begin{equation*}
\rho -p_{rad}=2\rho \geq 0
\end{equation*}%
\begin{equation*}
\rho -p_{\perp }=\frac{b^{2}}{4\pi }\left( -2+\sqrt{\left( \overline{Y}%
\right) ^{-4}+1}+\frac{1}{\sqrt{\left( \overline{Y}\right) ^{-4}+1}}\right) =%
\frac{b^{2}}{4\pi }\left( -2+\frac{\left( \overline{Y}\right) ^{-4}+2}{\sqrt{%
\left( \overline{Y}\right) ^{-4}+1}}\right) \geq 0
\end{equation*}%
the WEC\ is satisfied.

c) The strong energy condition (SEC):%
\begin{equation*}
\text{ requires}\ :\ \ \rho +\sum p_{k}=\frac{b^{2}}{2\pi }\left( 1-\frac{1}{%
\sqrt{\left( \overline{Y}\right) ^{-4}+1}}\right) \geq 0
\end{equation*}%
and defines the sign of the acceleration due to gravity and is fulfilled in
our case.

Notice because all the energy conditions are satisfied, not exotic
matter-energy need to be introduced in order to explain any anomalous
behaviour of the fields. Also is some references was claimed about that the
metric coefficient take a de Sitter behaviour when $r\rightarrow 0$. And
this fact is obvious: all regular solution necessarily has this form $%
1-Ar^{2}$ (A: some constant factor) in order to avoid \ divergences produced
by terms as $\frac{1}{r^{n}}$ with n\TEXTsymbol{>}1 near the origin.

\section{Horizons: physical considerations}

The metric, as was shown in $^{1}$, is absolutely regular\footnote[2]{%
However, as was shown in ref.[1], the global properties of the spacetime are
easily seen writting the line element (1) as a function of the $Y\left(
r\right) :$ $ds^{2}=-e^{2\Lambda }dt^{2}+e^{-2\Lambda }dY^{2}+Y^{2}\left(
d\theta ^{2}+\sin ^{2}\theta \,d\varphi ^{2}\right) $} being in our case, a
static spherically symmetric spacetime (SSS) with $g_{tt}\neq g_{rr}^{-1}$.
This fact make that the number of horizons in $g_{tt}$ can be none, one
(extreme) or two depending on the specific value of the parameters of the
solution, but $g_{rr}$ present only one extreme horizon (it is $g_{rr}=0$).
The localization of the minimum of $-g_{tt}$ coincides with the localization
of the point $g_{rr}=0$.

The physical interpretation is: because the form of the spacetime depends on
the distribution of mass/energy, that in the point of maximum value of the
electric field $g_{rr}$ takes the minimum value possible that correspond
with a spherical surface the radius $r_{cr}=-\frac{\sqrt[3]{2}}{a}r_{o}$ (at
this point we are over the surface/border of the electric spherical
configuration ). However, this fact don't disturb the computation of the
electromagnetic mass where the integration of the $T_{0}^{0}$ component (the
energy) was made, then the identification between the electromagnetic and
gravitational mass remains unaltered. Then, the r coordinate is well behaved
wherever be the behaviour of $g_{tt}$(one, two or none horizons) because
remains $g_{rr}\geq 0$ in all the manifold. When $g_{tt}$ present 2 horizons
there exists one region where the manifold takes Euclidean signature in the
coordinate system the field loses its electric character into this region).
The important thing for the analysis of this Section is that $%
g_{tt}=e^{2\Lambda }$ given by (2) is finite and regular without divergences
or incompleteness of any type in all the manifold and $g_{rr}$ presents a
regular horizon at the radius corresponding to minimum vale of $\overline{Y}%
^{2}$.

\subparagraph{Line element}

Now, we introduce in the general line element (1) a set of null modified
coordinates suitable to describe the characteristic null surface at the
extreme horizon $r_{extr}\left( Y_{extr}=Y\left( r_{extr}\right) \right) $
in $g_{rr}$ (angular part remaining the same) 
\begin{equation}
\frac{du+dv}{2}=dt\ \ \ \ \ \ ,\ \ \ \ \frac{dv-du}{2}=2\left( f^{\prime
}\left( r\right) +1\right) dr\   \tag{6}
\end{equation}%
\begin{equation*}
\ \ f^{\prime }\left( r\right) =\frac{\left[ \ \left( a\left\vert
r\right\vert \right) ^{n}+\left( mn-1\right) r_{0}\right] \left[ 1-\left(
a\left\vert r\right\vert \right) ^{-n}r_{0}\right] ^{m}}{\left( a\left\vert
r\right\vert \right) ^{n}-r_{0}}-1
\end{equation*}%
\begin{eqnarray*}
ds^{2} &=&-\frac{e^{2\Lambda }}{4}\left( du+dv\right) ^{2}+\frac{%
e^{-2\Lambda }}{4}\left( -du+dv\right) ^{2} \\
&=&-Sinh\left( 2\Lambda \right) \frac{\left( du^{2}+dv^{2}\right) }{2}%
-Cosh\left( 2\Lambda \right) dudv
\end{eqnarray*}%
In the coordinates (6) and with the values of the parameters fixed to $%
a=-0,9 $ and $m=1$ and $n=3,$ the line element is manifestly regular
everywhere, presenting a good behaviour also for $r\rightarrow 0$ and $%
r\rightarrow \infty $ where the spacetime is flat ($\Lambda \rightarrow 0$).
Also similar happens in the Schwarzschild line element when Kruskal type
coordinates are introduced in $r=2m$, but in our case this is not a maximal
analytical extension because this SSS is itself complete and need not be
extended..

\subparagraph{Geodesic completeness}

>From the geodesic point of view and for a static spherically symmetric
spacetime (SSS), the study of the radial case is sufficient to analyze its
completeness . Then, explicit computation of the radial geodesics lead to$%
^{9,13,15}$:%
\begin{equation}
\frac{dt}{ds}=e^{-2\Lambda }C  \tag{7}
\end{equation}%
\begin{equation*}
\frac{dr}{ds}=\pm \left( \delta e^{-2\Phi }+e^{-2\left( \Lambda +\Phi
\right) }C^{2}\right) ^{1/2}
\end{equation*}%
\begin{equation*}
\frac{dr}{dt}=\pm \frac{e^{2\Lambda }}{C}\left( \delta e^{-2\Phi
}+e^{-2\left( \Lambda +\Phi \right) }C^{2}\right) ^{1/2}
\end{equation*}%
where as in ref.$^{9}$ the constant C characterizes the test particle under
consideration, and $C>1$, $=1$, $<1$\ correspond respectively to finite,
zero and imaginary velocity of the particle at infinity (an infinite value
of C represents a light ray $ds=0$); and the constant $\delta =1,0$ for
timelike or null geodesics respectively. From above equations is easily
seen, that all non-spacelike geodesics can be extended to arbitrary values
of the affine parameter:

i) Radial null geodesics:

from equations (7) with $\delta =0$, we see that: $0\leq \left\vert \frac{dt%
}{ds}\right\vert =\left\vert \frac{dr}{ds}e^{-\Lambda }\right\vert \leq
\left\vert C\right\vert $, from where it follows that these geodesics are
complete.

ii)Radial timelike geodesics:

from equations (7) with $\delta =1$, we see that: $0\leq \left\vert \frac{dt%
}{ds}\right\vert =\left\vert e^{2\Lambda }+\left( \frac{dt}{ds}\right)
^{2}e^{4\Lambda }\right\vert ^{1/2}\leq \left\vert e^{2\Lambda
}+C^{2}\right\vert ^{1/2}$, from where it follows that these geodesics are
complete.

\subparagraph{Curvature}

As was pointed out in ref.$^{1}$ and for the set of parameters fixed to the
values: $a=-0,9$ and $m=1$ and $n=3,$ the explicit computation of the
curvature scalar indicate us that the spacetime is singularity free
(geometrodynamics units $^{1}$):%
\begin{equation*}
R_{a}^{a}=R=2(G_{00}-G_{22})=\left( -2+\frac{\overline{Y}^{-4}+2}{\sqrt{%
\overline{Y}^{-4}+1}}\right)
\end{equation*}%
that clearly shows that $R\rightarrow 0$ when $r\rightarrow 0,\infty $ and
has not divergences in all the manifold\footnote[3]{%
Moreover, since that the tensor $R_{bcd}^{a}$ for the spherically symmetric
static metric under consideration here is pairwise diagonal, the Kretschmann
scalar $\mathcal{K}$ is a sum of squares:%
\begin{equation*}
\mathcal{K=}R_{\ \ \ cd}^{ab}R_{\ \ \ ab}^{cd}=4\left( R^{0}\,_{110}\right)
^{2}+8\left( R^{2}\,_{112}\right) ^{2}+8\left( R^{0}\,_{220}\right)
^{2}+4\left( R^{3}\,_{223}\right) ^{2}\,\,
\end{equation*}%
where the components of the Riemann tensor were explicit computed in I,
that, with the corresponding values for the metric coefficients of the
solution obtained in our previous work, indicate us that the metric presents
no problems also$.$}.

\subparagraph{Extreme horizon localization}

In order to determine the horizon in the extreme case for $g_{tt}$ (because
we have been seen in our previous paper that we can have none, one or two
horizons for $g_{tt}$), \ the $b$ value must be fixed at the last stage of
the computation. As we pointed out before, $g_{rr}=0$ at some critical
radius where the electric field take its maximum value and $g_{tt}$ presents
the maximum curvature. In the cases where there exist an extreme horizon of
this type, as in the case presented here, the position of it can be find as
follows:

i) select a set of parameters $a,m,n,r_{0}$, e.g. $a,1,3,r_{0}\rightarrow
Y^{2}\equiv \left[ 1-\left( \frac{r_{o}}{a\left\vert r\right\vert }\right)
^{3}\right] ^{2}r^{2}$

ii) compute where is the extreme of the electric field, the metric or where $%
g_{rr}=0,$ e.g: $\frac{dF_{10}}{dr}=0$ $\rightarrow -\frac{1}{2}=\left( 
\frac{r_{o}}{a\left\vert r\right\vert }\right) ^{3}$\newline

iii) from the previous point, determine the critical values $r_{cr}$ (or $%
Y_{cr}\left( r\right) $)$\rightarrow r_{cr}=-\frac{\sqrt[3]{2}}{a}r_{o}\ \
\left( Y_{cr}=-3\frac{\sqrt[3]{2}}{2}\frac{r_{o}}{a}\right) $. For $a=-0.9$
and $r_{0}=1$(our case): $r_{cr}=1.39991\left( Y_{cr}=2.09987\right) $

iv) with the critical values $r_{cr}$ (or $Y_{cr}\left( r\right) $) in the
expression for $g_{tt}=0$ we determine the corresponding $b$ value.

Then for the determined $b$ value the extreme horizon in $g_{tt}$ is in $%
r=r_{cr}$ (Fig.1).

In resume, we saw that the metric coefficients are not reciprocals
(different) because the line element is the more general spherically
symmetric adecuated to the symmetries of the Born-Infeld theory. This fact
is based in the observation that the Born-Infeld energy-momentum tensor
takes the same form in the tetrad defined for expressions in I \ that an
anisotropic fluid then, it defines that the correct ansatz to solve the
problem$^{15}$ .

The spacetime is geodesically complete and singularity free: all
non-spacelike geodesics can be extended to arbitrary values of the affine
parameter and the curvature (and Kretschmann) scalar presents no
divergencies for all values of r (see ref.$^{9})$.

\subsection{The absolute b field, the electron mass and the cosmological
constant}

In this Section we will to consider some physical aspects of some quantities
naturally introduced in the BI theory: the absolute field $b$ and the\
radius $r_{0}$ that was related in the earliest references$^{2,3,4}$ with
the electron radius$.$

>From the mass formula (2.2) and having account that $b^{2}r_{o}^{4}=Q^{2}$
we have an expression for $r_{0}$ as a function of the mass and charge of
the electron: 
\begin{equation*}
r_{0}\cong 3.4831.10^{-13}cm
\end{equation*}%
that is a more accurate value for the electron radius than the given in
reference$^{2}$, or for logaritmic type Lagrangians as in$^{19}$. Now, the
value of the absolute field $b$ is easily determined%
\begin{equation*}
b\cong 3.96718.10^{15}esu/cm^{2}
\end{equation*}%
However, the enormous magnitude of this absolute $b$ field justifies the
application of the Maxwell's equations in their classical form in all cases,
except those were the inner structure of the electron is concerned (fields
of the order $b$, distance or wavelength of order $r_{0}$.

Is interesting to note that if we take the value of $%
b_{cr}=3,56647.10^{24}esu/cm^{2}$(the value of $b$ where $g_{tt}$ have the
extreme horizon) and $Q\simeq e/3$ (as suggested time ago for Rosen as for
the quarks case) , the obtained value for $r_{0}$ is $6,7.10^{-18}cm$ given
a mass of the order of $9,5.10^{-23}g\sim 5,33.10^{4}GeV.$

Finishing this discussion about orders of magnitude and physical
considerations; in the reference$^{10}$ a new non-abelian generalization of
the BI theory was proposed, from the first principles, with many interesting
properties as to be absolutely independent of the symmetry gauge groups,
then of the trace prescriptions$^{11}$ and presents a full generalized
non-linear duality$^{12}$. A new wormhole solution was obtained solving the
Einstein-NABI equations leading to the following ordinary differential
equation for the scale factor $a$ 
\begin{equation*}
3\left[ \left( \frac{\overset{.}{a}}{a}\right) ^{2}-\frac{1}{a^{2}}\right]
=2G\left( b^{2}-4\pi \Lambda \right) -2Gb^{2}\left[ 1+6\left( \frac{r_{0}}{a}%
\right) ^{4}\left( 1+\left( \frac{r_{0}}{a}\right) ^{4}\right) \right] ^{1/2}
\end{equation*}%
As was shown in$^{11}$ the integrability condition for this equation is
(geometrodynamic units$^{10}$)%
\begin{equation*}
\left( b^{2}=4\pi \Lambda \right)
\end{equation*}%
this fact constrain the value of for the non-abelian case to $b\sim
1.23506.10^{58}esu/cm^{2}$ and $r_{0}\sim l_{Planck}$. It is obvious that we
can consider, if the value of $b$ is of the above order, an effective
cosmological constant $\Lambda _{eff}\sim \left( b^{2}-4\pi \Lambda \right)
\sim 0$ where the big value of $\Lambda _{Planck}$ is screened by an
absolute field associated with a generalized NABI theory. This fact can be
of great importance in models based in d-brane/superstring\ theories where
the b-field is naturally related with the d-brane (string) tension$^{11}$.

\section{\protect\bigskip Regularity, inconsistencies and no-go theorems}

As was clearly explained in reference$^{1}$, and we have been mention in the
Introduction, the main point in order to obtain SSS in the (nonlinear) EBI
theory are the following

i) Most general spherically symmetric line element as starting point,
compatible with the symmetry of the an anisotropic fluid (the Born-Infeld
energy-momentum tensor has this form), that lead a general solution as
expression (1)

ii) strict physical requirements of regularity for the electric field:

The electric field cannot be greater that the absolute field $b$ 
\begin{equation*}
\left. F_{01}\right\vert _{r=r_{o}}<b
\end{equation*}%
The electric field need to be zero at the origin\footnote[4]{%
r corresponding to a coordinate basis} $r\rightarrow 0$%
\begin{equation*}
\left. F_{01}\right\vert _{r=0}=0
\end{equation*}%
and for r$\rightarrow \infty $ Coulombian behaviour%
\begin{equation*}
\,\ \ \ \ \ \ \ \ \ \ \ \ \ \ \ \ \ \ \ \ \ \ \ \ \ \ \left.
F_{01}\right\vert _{r\rightarrow \infty }=0\,\ \ \ \ \ \ \ \ \ \ \ \ \text{%
asymptotically \ Coulomb}
\end{equation*}%
The Lagrangian of the regular electric monopole of$^{1}$ is 
\begin{equation*}
L=\frac{b^{2}}{4\pi }\left( 1-\sqrt{\left[ 1-\left( \frac{F_{01}}{b}\right)
^{2}\right] }\right)
\end{equation*}%
where 
\begin{equation*}
\left( \frac{F_{01}}{b}\right) ^{2}=\frac{1}{\overline{Y}^{4}+1}
\end{equation*}%
never diverges having account the behaviour of$\ Y^{2}\left( r\right) $ for
the selected values for the parameters [see (4)]%
\begin{eqnarray*}
\text{for\ }a &=&-0,9\text{, }m=1\text{ and }n=3\text{ \ \ \ \ \ \ \ } \\
\text{\ when\ }r &\rightarrow &0,\ Y^{2}\left( r\right) \rightarrow \infty ,%
\text{ \ \ \ and when \ }r\rightarrow \infty ,\ Y^{2}\left( r\right)
\rightarrow r^{2}
\end{eqnarray*}%
then%
\begin{equation*}
\text{when\ }r\rightarrow 0,\ \left( \frac{F_{01}}{b}\right) ^{2}\rightarrow
0,\text{ and when \ }r\rightarrow \infty ,\ \left( \frac{F_{01}}{b}\right)
^{2}\rightarrow 0
\end{equation*}%
and because the maximum value of the electric field never takes the absolute
field $b$ value e.g:$\left. F_{01}\right\vert _{r=r_{o}}<b\Rightarrow
L\rightarrow 0$ at r$\rightarrow 0$ that is the Maxwellian behaviour. Also:%
\begin{equation*}
\text{when\ }r\rightarrow 0,\text{ }\frac{dL}{dS}=\sqrt{\left[ 1-\left( 
\frac{F_{01}}{b}\right) ^{2}\right] }\rightarrow 1
\end{equation*}%
where: $S\equiv -\frac{1}{4}F_{\alpha \beta }F^{\alpha \beta }.$ Is easily
seen that $L$ as a function of $F$ has not branches. Because the
energy-momentum tensor is absolutely regular, as we show in expressions (5),
and goes to zero when\ $r\rightarrow 0$, the density $T_{00}$ goes obviously
to zero (Fig.2) according to physical requirements of regularity.
Schematically%
\begin{equation*}
F\rightarrow 0\Rightarrow \rho \rightarrow 0\Leftrightarrow g\rightarrow 1%
\text{(flat space,vacuum)}
\end{equation*}%
Notice that in particular SSS solutions from nonlinear electrodynamics$^{7}$
the points i) and ii) were not having account and, in order to avoid
divergences of $\frac{dL}{dS},$ a maximal density $\rho $ at the centre of
the spherical configuration was required.

To finish this Section, we show that the field 
\begin{equation*}
\mathbb{F}_{01}=\frac{F_{01}}{\sqrt{1-\left( \overline{F}_{01}\right) ^{2}}}
\end{equation*}%
that is related with the D-field of \ the an electrodynamics in a continuum
medium$^{20}$ has similar regular well behaviour that the electric field as
is easily seen from above equation (Figure 3). As far as we know, all the
other solutions coming from NED\ formulations$^{6,7,8}$ present divergencies
of this D-field $\left( \mathbb{F}_{01}\right) $ when $r\rightarrow 0$.

\section{Dyon: statement of the problem}

Is clearly important in order to finish our research, to consider the more
general case that is the static spherically symmetric solution (SSS) in EBI\
theory with electric and magnetic field: the dyon. To do this, we propose
the following line element for the static Born-Infeld monopole as in$^{1}$

\begin{equation}
ds^{2}=-e^{2\Lambda }dt^{2}+e^{2\Phi }dr^{2}+e^{2F\left( r\right) }d\theta
^{2}+e^{2G\left( r\right) }\sin ^{2}\theta \,d\varphi ^{2}  \tag{8}
\end{equation}%
where the components of the metric tensor are 
\begin{equation}
\begin{array}{cccc}
g_{tt}=-e^{2\Lambda } &  &  & g^{tt}=-e^{-2\Lambda } \\ 
g_{rr}=e^{2\Phi } &  &  & g^{rr}=e^{-2\Phi } \\ 
g_{\theta \theta }=e^{2F} &  &  & g^{\theta \theta }=e^{-2F} \\ 
g_{\varphi \varphi }=\sin ^{2}\theta \,e^{2G} &  &  & g^{\varphi \varphi }=%
\frac{e^{-2G}}{\sin ^{2}\theta }%
\end{array}
\tag{9}
\end{equation}%
For the obtention of \ the Einstein-Born-Infeld equations system we use the
Cartan's structure equations method$^{1}$, that is most powerful and direct
where we work with differential forms and in a orthonormal frame (tetrad).
The line element (7) in the 1-forms basis takes the following form 
\begin{equation}
ds^{2}=-\left( \omega ^{0}\right) ^{2}+\left( \omega ^{1}\right) ^{2}+\left(
\omega ^{2}\right) ^{2}+\left( \omega ^{3}\right) ^{2}  \tag{10}
\end{equation}%
were the forms are 
\begin{equation}
\begin{array}{cccc}
\omega ^{0}=e^{\Lambda }dt &  & \Rightarrow & dt=e^{-\Lambda }\omega ^{0} \\ 
\omega ^{1}=e^{\Phi }dr &  & \Rightarrow & dr=e^{-\Phi }\omega ^{1} \\ 
\omega ^{2}=e^{F\left( r\right) }d\theta &  & \Rightarrow & d\theta
=e^{-F\left( r\right) }\omega ^{2} \\ 
\omega ^{3}=e^{G\left( r\right) }\sin \theta \,d\varphi &  & \Rightarrow & 
d\varphi =e^{-G\left( r\right) }\left( \sin \theta \right) ^{-1}\omega ^{3}%
\end{array}
\tag{11}
\end{equation}%
Now, following the standard procedure of the structure equations (Appendix
in ref.$^{1}$) for to obtain easily the components of the Riemann tensor, we
can construct the Einstein equations 
\begin{equation}
G^{1}\,_{2}=-e^{-\left( F+G\right) }\frac{\cos \theta }{\sin \theta }%
\partial _{r}\left( G-F\right)  \tag{12}
\end{equation}

\begin{equation}
G^{0}\,_{0}=e^{-2\Phi }\Psi -e^{-2F}  \tag{13}
\end{equation}%
\begin{equation*}
\Psi \equiv \left[ \partial _{r}\partial _{r}\left( F+G\right) -\partial
_{r}\Phi \,\partial _{r}\left( F+G\right) +\left( \partial _{r}F\right)
^{2}+\left( \partial _{r}G\right) ^{2}+\partial _{r}F\,\partial _{r}G\right]
\end{equation*}

\begin{equation}
G^{1}\,_{1}=e^{-2\Phi }\left[ \partial _{r}\Lambda \,\partial _{r}\left(
F+G\right) +\partial _{r}F\,\partial _{r}G\right] -e^{-2F}  \tag{14}
\end{equation}

\begin{equation}
G^{2}\,_{2}=e^{-2\Phi }\left[ \partial _{r}\partial _{r}\left( \Lambda
+G\right) -\partial _{r}\Phi \,\partial _{r}\left( \Lambda +G\right) +\left(
\partial _{r}\Lambda \right) ^{2}+\left( \partial _{r}G\right) ^{2}+\partial
_{r}\Lambda \,\partial _{r}G\right]  \tag{15}
\end{equation}

\begin{equation}
G^{3}\,_{3}=e^{-2\Phi }\left[ \partial _{r}\partial _{r}\left( F+\Lambda
\right) -\partial _{r}\Phi \,\partial _{r}\left( F+\Lambda \right) +\left(
\partial _{r}\Lambda \right) ^{2}+\left( \partial _{r}F\right) ^{2}+\partial
_{r}F\,\partial _{r}\Lambda \right]  \tag{16}
\end{equation}

\begin{equation}
G^{1}\,_{3}=G^{2}\,_{3}=G^{0}\,_{3}=G^{0}\,_{2}=G^{0}\,_{1}=0  \tag{17}
\end{equation}

In the tetrad defined by (11), the energy-momentum tensor of Born-Infeld
takes a diagonal form, being its components the following 
\begin{equation}
-T_{00}=T_{11}=\frac{b^{2}}{4\pi }\left( \frac{\mathbb{R}-1}{\mathbb{R}}%
\right)  \tag{18}
\end{equation}%
\begin{equation}
T_{22}=T_{33}=\frac{b^{2}}{4\pi }\left( 1-\mathbb{R}\right)  \tag{19}
\end{equation}%
where for the dyon 
\begin{equation}
\mathbb{R}\equiv \sqrt{\left[ 1-\left( \frac{F_{01}}{b}\right) ^{2}\right] %
\left[ 1-\left( \frac{F_{32}}{b}\right) ^{2}\right] }  \tag{20}
\end{equation}%
of this manner, one can see from the Einstein equation (12) the
characteristic property of the spherically symmetric space-times$^{15}$%
\begin{equation}
G^{1}\,_{2}=-e^{-\left( F+G\right) }\frac{\cos \theta }{\sin \theta }%
\partial _{r}\left( G-F\right) =0\,\ \ \ \ \Rightarrow \,\ \ \ G=F  \tag{21}
\end{equation}%
Notice for that the interval be a spherically symmetric one, the functions $%
F\left( r\right) $ and $G\left( r\right) $ must be equal. As we saw in the
precedent paragraph the components of the energy-momentum tensor of BI
assures this condition in a natural form. Also it is interesting to see from
eqs. (18) and (19) that the energy-momentum tensor of Born-Infeld has the
same form as the energy-momentum tensor of an anisotropic fluid$^{15}$.

\section{Dyon: field equations vs. energy-momentum conservation}

To obtain explicitly the electromagnetic fields in the tetrad (11), the
following equations need to be solved%
\begin{equation*}
dF=0\text{ \ \ \ ; \ \ \ \ \ }d\widetilde{\mathbb{F}}=0
\end{equation*}%
the Bianchi identity and the equations of motion for the electromagnetic
field, that in the same language that in$^{1}$, are 
\begin{equation}
\nabla _{a}\mathbb{F}^{ab}=\nabla _{a}\left[ \frac{F^{ab}}{\mathbb{R}}+\frac{%
P}{b^{2}\mathbb{R}}\widetilde{F}^{ab}\right] =0\,\ \ \ \ \ \ \ \ \ \ \ \ \ \
\ \ \ \ \ \ \ \ \left( field\,equations\,\right)  \tag{22}
\end{equation}%
\begin{equation}
\nabla _{a}\,\,\widetilde{F}^{ab}=0\,\ \ \ \ \ \ \ \ \ \ \ \ \ \ \ \ \ \ \ \
\ \ \ \ \ \ \ \ \ (\ Bianchi^{\prime }s\,\ identity)\ \ \ \ \ \ \ \ \ \ \ \
\ \ \ \ \ \ \ \ \ \ \ \ \ \ \ \ \ \ \ \ \ \   \tag{23}
\end{equation}%
where 
\begin{equation}
P\equiv -\frac{1}{4}F_{\alpha \beta }\widetilde{F}^{\alpha \beta }  \tag{24}
\end{equation}%
\begin{equation}
S\equiv -\frac{1}{4}F_{\alpha \beta }F^{\alpha \beta }  \tag{25}
\end{equation}%
\begin{equation}
\mathbb{R}\equiv \sqrt{1-\frac{2S}{b^{2}}-\left( \frac{P}{b^{2}}\right) ^{2}}%
=\sqrt{\left[ 1-\left( \frac{F_{01}}{b}\right) ^{2}\right] \left[ 1+\left( 
\frac{F_{32}}{b}\right) ^{2}\right] }  \tag{26}
\end{equation}%
The above equations can be solved explicitly giving the follow result%
\begin{equation*}
dF=0\text{ }\Rightarrow F_{01}e^{\Lambda +\Phi }=A\left( r\right) \text{ \ \
and \ \ }F_{23}e^{2G}=B
\end{equation*}%
and%
\begin{equation}
d\widetilde{\mathbb{F}}=0\Rightarrow \mathbb{F}_{23}=F_{23}\sqrt{\frac{%
1-\left( \frac{F_{01}}{b}\right) ^{2}}{1+\left( \frac{F_{32}}{b}\right) ^{2}}%
}=C\left( r\right) e^{-\left( \Lambda +\Phi \right) }\text{ \ \ and} 
\tag{26.1}
\end{equation}%
\begin{equation}
F_{01}\sqrt{\frac{1+\left( \frac{F_{32}}{b}\right) ^{2}}{1-\left( \frac{%
F_{01}}{b}\right) ^{2}}}=De^{-2G}  \tag{26.2}
\end{equation}%
>From the above to equations is easily seen that 
\begin{equation}
u\equiv \sqrt{\frac{1+\left( \frac{F_{32}}{b}\right) ^{2}}{1-\left( \frac{%
F_{01}}{b}\right) ^{2}}}=\frac{D}{A\left( r\right) }=\frac{B}{C\left(
r\right) }  \tag{27}
\end{equation}%
then, without lost generality, we can make $B=D$ and $A\left( r\right)
=C\left( r\right) $.

Now is very early to say something about the constants, then, the real form
of the electromagnetic field. In order to corroborate and restrict the kind
of solutions we must go to the equations of the energy-momentum conservation
in the tetrad (11) 
\begin{equation}
\nabla _{a}T^{ab}=0  \tag{28}
\end{equation}%
\begin{equation*}
\nabla _{1}T^{11}=0\rightarrow \partial _{r}T^{11}+2\partial _{r}G\left(
T^{11}-T^{22}\right) =0
\end{equation*}%
\begin{equation*}
\nabla _{1}T^{11}\equiv 0,\ \ \ \ \ \ \ \ \ \nabla _{3}T^{33}=\nabla
_{2}T^{22}=0
\end{equation*}%
where the non vanishing connection coefficients from the tetrad defined by
(11) are easily obtained from $\Gamma _{cd}^{b}=\left\langle \omega
^{b},\nabla _{d}E_{c}\right\rangle $ as usual.

Because the components of the energy-momentum tensor in the tetrad (11) can
be explicitly written as functions of the $u$ invariant (27)%
\begin{equation}
-T_{00}=T_{11}=\frac{b^{2}}{4\pi }\left( 1-u\right) ,\ \ \ \ \ \ \
T_{22}=T_{33}=\frac{b^{2}}{4\pi }\left( 1-u^{-1}\right)  \tag{29}
\end{equation}%
the final form of $u$ is immediately determined%
\begin{equation}
u=\sqrt{\left( \frac{r_{0}}{e^{G}}\right) ^{4}+1}  \tag{30}
\end{equation}%
that is the expected result with $r_{0}$ the integration constant with
length units that was related with the radius of the electron in ref.$^{2}$.
Notice that from the point of view of the energy-momentum conservation only
the combination of the electromagnetic invariants $u$ is determined. Also,
was pointed out in$^{2}$, this is a direct consequence of the Von Laue's
theorem in an \textit{unitarian electromagnetic theory} as is the
Born-Infeld case.

Now, from eqs.(27), (26.1) and (26.2) and considering $F_{23}=De^{-2G}$
obtained from the field equations(notice an scale arbitrarity in the radius $%
D/A\left( r\right) $) we arrive to%
\begin{equation}
F_{01}=b\sqrt{\frac{1-\left( D^{2}/b^{2}r_{0}^{4}\right) }{1+\left(
e^{4G}/r_{0}^{4}\right) }}  \tag{31}
\end{equation}%
that is also an expected result considering that one recover the expression
for the electric monopole putting $D=0$. In the same manner that for the
pure electric case we can suppose $D=Q_{m}$ and $Q_{T}=br_{0}^{2}$(because
the units allows us). Then%
\begin{equation}
F_{01}=b\sqrt{\frac{Q_{T}^{2}-Q_{M}^{2}}{Q_{T}^{2}+\left( b^{2}e^{4G}\right) 
}}\text{ \ and \ \ \ \ }F_{23}=e^{-2G}Q_{m}  \tag{32}
\end{equation}%
Is important to observe that the concept of charge is only an asymptotic
idea. The last expressions of the electromagnetic fields as functions of
charges (that doesn't exist! they are there due by the nonlinearity of the
theory) is only a "far away" hint in order to interpret this unitarian
theory as in the maxwellian (linear and dualistic) case. In resume:

i) for the dyonic solution the metric will have the same behaviour and
properties that the electric case previously studied by the authors in
reference$^{1}$: is obvious that the Einstein equations remain the same
although the \ presence of the electric and magnetic fields.

ii) because the concept of charge is connected with the nonlinear behaviour
of the electromagnetic fields, the same electric and magnetic components are
interrelated as is easy to see in $F_{01}=b\sqrt{\frac{1-\left(
D^{2}/b^{2}r_{0}^{4}\right) }{1+\left( e^{4G}/r_{0}^{4}\right) }}$ that
depends on the magnetic part due the constant $D=Q_{m}.$

iii) from ii) is obvious that the one must necessarily forget the idea of
charge as $Q_{m},Q_{T}$ etc being the correct expression for the fields%
\begin{equation}
F_{01}=b\sqrt{\frac{1-\left( D^{2}/b^{2}r_{0}^{4}\right) }{1+\left(
e^{4G}/r_{0}^{4}\right) }}\text{ \ \ \ \ \ and \ }F_{23}=e^{-2G}D\text{\ \ }
\tag{33}
\end{equation}

\section{Dyon solution: the electromagnetic field}

As was pointed out previously, the energy momentum tensor only take account
on the electromagnetic fields content, e.g. $u$ (27), through $r_{0}$ then,
the Einstein equations for the electric, magnetic and dyonic cases are
formally the same expressions. However, the interpretation of the charges,
the fields and maximum values will be different in each case.

Regarding the electric and magnetic cases (e.g. Figure 1 of $^{1}$ and
Figure 4 in this paper) there are regular with coulombian behaviour at $%
r\rightarrow \infty $ and going to $\rightarrow 0$ when $r\rightarrow 0$
presenting the maximum value at the spherical surface with $r_{B}=2^{1/3}%
\frac{r_{0}}{\left\vert a\right\vert }.$

The other important point is that the $\mathbb{F}_{ab}$ field (called in
some references as D-field$^{6,7,8}$) is absolutely regular in $r\rightarrow
0$ \ (Figure 3) presenting an analog behaviour as $F_{ab}$, remaining under
the maximum value $b$ in all the spacetime. This point is important because,
as far as we known, all the solutions of nonlinear electrodynamics present a
divergence of $\mathbb{F}_{ab}$ and also in the magnetic case$^{5,6,7}$, and
many speculations were given about the impossibility of simultaneous
regularity of $\mathbb{F}$ and $F$. But the problem is solved having account
the conditions pointed out previously in Section 4.

The dyonic case has a mixing between the pure electric and magnetic
situation. The total intensity attributed to the spherical configuration is
distributed between the electric and magnetic fields as dictated by
equations (33). The electric field depends on the value of the magnetic
field (this is the reason because the meaning of the word "charge" in this
case is obscure). Also in this case the electric, magnetic and $\mathbb{F}$
field are absolute regular at the origin and remaining all of them bounded
by below the b value.

\section{Discussion}

Contrarily to the claimed in ref.$^{8}$ we shown a SSS solution of the
Born-Infeld theory that have regular behaviour without divergencies of the $%
\mathbb{F}$ field at r=0 and cusps or branches in the Lagrangian. The main
reason because the circumvention of these claims$^{6,7,8}$ is possible, is
that we start from a more general SSS line element with $g_{tt}\neq
g_{rr}^{-1}$ and strict regularity conditions over the electromagnetic field 
\begin{equation*}
\left. F\right\vert _{r=r_{o}}<b
\end{equation*}%
\begin{equation*}
\left. F\right\vert _{r=0}=0
\end{equation*}%
this fact permit to introduce the correct regularity conditions over the
fields of the theory given the good features described in our previous work$%
^{1}$ and the analyzed here:

i) the field $\mathbb{F\equiv }\frac{\mathbb{\partial }L}{\partial F}$
(D-field in ) has regular behavior in $r\rightarrow 0$ (Figure 3)

ii) the metric is continuous at the border of the spherical configuration $%
r_{B}$: there are not "inner and outer" matching conditions in this limit
for the solution, only one continuous function (i.e.Fig.1).

iii) from the point of view of the geodesics: the spacetime is geodesically
complete and not continuation is required.

iv) the extreme horizon $g_{rr}=0$ is absolutely regular, being a null
surface precisely the sphere $r_{B}=2^{1/3}\frac{r_{0}}{\left\vert
a\right\vert }$, as is easily seen when in the line element modified null
coordinates are specially introduced.

v) The physical interpretation is, because the form of the spacetime depends
on the distribution of mass/energy, in the point of maximum value of the
electric field $g_{rr}$ takes the minimum value physically possible that
correspond with an spherical surface the radius $r_{cr}=-\frac{\sqrt[3]{2}}{a%
}r_{o}$ (at this point we are over the surface/border of the spherical
configuration of the electric field)

vi) the solution with the integration constant $M\neq 0$ is still regular
due the behaviour of the $Y\left( r\right) $ with $-1<a<0\ \left( e.g.:a=-0,9%
\text{; }m=1\text{ and }n=3\right) .$ However with M$\neq 0$,

$\circ $ we cannot identify the electromagnetic with the gravitational mass;

$\circ $ the solution is not invariant under the change $r\leftrightarrow
-r; $

$\circ $ when $r_{o}\rightarrow 0$ (Maxwell/Reissner-Nordstr\"{o}m limit)
the regularity is obviously broken.

vii) the dyonic case is analog to the electric and magnetic cases, but the
electric field depends on the magnetic field (as dictated by equations (33))
being, in some sense, modulated by it. The total intensity attributed to the
dyonic spherical configuration is distributed between the electric and
magnetic fields. The metric have the same regular shape in the three
situations: the energy-momentum tensor (then the Einstein equations (12-17))
don't see the specific structure of the magnetic and electric components of
the $F_{ab}$: the specific structure of the electromagnetic fields are
determined by the dynamical field equations plus the conservation of the
energy momentum tensor, as was shown in Section 6.

viii) the BI theory is \textit{unitarian}: the concept of charge is
asymptotically attributed due the non-linear behaviour of the
electromagnetic field: the electromagnetic field is the basic identity.
Then, the proposal of a "sum rule" of\text{ }$Q_{e}^{2}$\text{and }$%
Q_{m}^{2} $ due to duality prescriptions coming from the Maxwell (linear)
theory is only an approximation, and we prefer present the fields with the
correct constants as were obtained. The full symmetries of the BI\ field
equations that give a more accurate answer to this point of general duality
and charges was investigated in ref.$^{12}$ by the authors, and will be
analyzed elsewhere$^{14}$.

\subsection{References}

\bigskip

$^{1}$D. J. Cirilo Lombardo, Journal of Math. Phys. 46, 042501 (2005)]

$^{2}$ M. Born and L. Infeld, Proc. Roy. Soc.(London) \textbf{144}, 425
(1934).

$^{3}$ B. Hoffmann, Phys. Rev. \textbf{47}, 887 (1935).

$^{4}$B. Hoffmann and L. Infeld, Phys. Rev. \textbf{51}, 765 (1937).

$^{5}$ M. Demianski, Found. Phys. Vol. 16 , No. 2, 187 (1986).

$^{6}$S. R. Hildebrandt and A. Ya. Burinskii, Phys. Rev. D \textbf{65},
104017 (2002), and references therein.

$^{7}$I. Dymnikova, Class. Quantum Grav. 21, 4417 (2004), and refernces
therein.

$^{8}$K. A. Bronnikov, Phys. Rev. Lett. \textbf{85}, 4641 (2000).

$^{9}$F. J. Chinea et al., Phys. Rev. D \textbf{45}, 481 (1992).

$^{10}$ M. Born , Proc. Roy. Soc. (London) \textbf{143}, 411 (1934).

$^{11}$D. J. Cirilo Lombardo, Class. Quantum Grav. \textbf{22}, 4987 (2005).

$^{12}$D. J. Cirilo Lombardo, Journal of Math. Phys. \textbf{48}, 032301
(2007).

$^{13}$S. W. Hawking and G. F. R. Ellis, \textit{The Large Scale Structure
of the Spacetime}, (Cambridge University Press, Cambridge, England, 1973).

$^{14}$D. J. Cirilo Lombardo, work in preparation

$^{15}$ D. Kramer et al., \textit{Exact Solutions of Einstein's Field
Equations, }(Cambridge University Press, Cambridge,1980).

$^{16}$ A. S. Prudnikov, Yu. Brychov and O. Marichev, \textit{Integrals and
Series }(Gordon and Breach, New York, 1986).

$^{17}$ K. A. Bronnikov, B. E. Meierovich and E. R. Podolyak, JETP\ \textbf{%
95}, 392 (2002).

$^{18}$ D. J. Cirilo Lombardo, Preprint JINR-E2-2003-221.

$^{19}$ H. H. Soleng, Phys. Rev. D \textbf{52}, 6178 (1995).

$^{20}$ L. D. Landau and E. M. Lifshitz,\textit{\ Electrodinamica de los
Medios Continuos,} (Revert\'{e}, Buenos Aires, 1974) .

\ 

\bigskip

\bigskip

\bigskip

\bigskip

\end{document}